# Flux growth of $Sr_{n+1}Ir_nO_{3n+1}$ (n=1, 2, ∞) crystals

K. Manna[1†*], G. Aslan-Cansever[1], A. Maljuk[1], S. Wurmehl[1], S. Seiro[1], B. Büchner[1]

[1]*Institute for Solid State Research, Leibniz IFW Dresden, Helmholtzstr. 20, 01069 Dresden, Germany*

[†] Presently at: Max Planck Institute for Chemical Physics of Solids, 01187 Dresden, Germany.

*Corresponding author email: kaustuvmanna@gmail.com

**Abstract**

Single crystals of iridates are usually grown by a flux method well above the boiling point of the $SrCl_2$ solvent. This leads to non-equilibrium growth conditions and dramatically shortens the lifetime of expensive Pt crucibles. Here, we report the growth of $Sr_2IrO_4$, $Sr_3Ir_2O_7$ and $SrIrO_3$ single crystals in a reproducible way by using anhydrous $SrCl_2$ flux well below its boiling point. We show that the yield of the different phases strongly depends on the nutrient/solvent ratio for fixed soak temperature and cooling rate. Using this low-temperature growth approach generally leads to a lower temperature-independent contribution to the magnetic susceptibility than previously reported. Crystals of $SrIrO_3$ exhibit a paramagnetic behavior that can be remarkably well fitted with a Curie-Weiss law yielding physically reasonable parameters, in contrast to previous reports. Hence, reducing the soak temperature below the solvent boiling point not only provides more stable and controllable growth conditions in contrast to previously reported growth protocols, but also extends considerably the lifetime of expensive platinum crucibles and reduces the corrosion of heating and thermoelements of standard furnaces, thereby reducing growth costs.

**Introduction**

In the past decade iridates have attracted a strong interest due to the alluring perspective of novel electronic and/or topological orders emerging from the interplay of spin–orbit coupling and electron correlations. The possibility of novel and unique physics in the $Sr_{n+1}Ir_nO_{3n+1}$ ($n = 1$, 2, and ∞) family of materials has been the focus of an increasingly large number of publications [1-4]. Due to complex hybridization of the Ir $d$ states with O $p$ states, intensive investigations are performed to understand the evolution of associated electronic structure and magnetic properties. While $SrIrO_3$ is found to be paramagnetic, $Sr_2IrO_4$ and $Sr_3Ir_2O_7$ show weak ferromagnetic behavior with transition temperatures of 225-240 K and 285 K, respectively. Different reports show a wide spread of transition temperatures for $Sr_2IrO_4$, while the paramagnetic susceptibility of $SrIrO_3$ yielded unreasonably large values when fitted with a Curie-Weiss law. Both observations strongly indicates the influence of sample quality on the physical properties of these systems [5].

From the crystal growth point of view, these materials are always grown from a flux due to their incongruently melting behavior [6]. Although single crystals of $Sr_2IrO_4$ ($n=1$, "214" hereafter), $Sr_3Ir_2O_7$ ($n=2$, "327" hereafter) and $SrIrO_3$ ($n=∞$, "113" hereafter) have been reported to grow from a $SrCl_2$ flux in platinum crucibles [1, 2, 5, 7-9], experiments are difficult to reproduce owing to an incomplete description of the growth conditions in many cases. The nutrient/solvent ratio, one of the key growth parameters, is not mentioned in various reports [5, 7, 8]. Surprisingly, the soak temperatures used are often well above the boiling point of the $SrCl_2$ flux (1250 °C): soak temperatures as high as 1480 °C and 1470 °C were reported for the growth of 214, 327 and 113 phases [7, 8, 10], and 1300 °C for the growth of the 214 phase [1, 9]. Note that 214 and 113 compounds have been reported to decompose in air at 1445°C and 1205°C, respectively [6]. In a boiling flux, crystals must grow far away from quasi-equilibrium conditions, which may lead to crystal imperfections [9]. Also, the actual ratio of nutrient to solvent changes at high temperatures as solvent evaporates affecting the reproducibility of growth experiments. At such elevated temperatures, the evaporated flux can severely attack heating and thermoelements of standard furnaces. In addition, heating Pt in air to 1300°C leads to a loss of 0.2% weight in 24 hours [11] due to formation of volatile $PtO_2$, and more importantly, the melt can become extremely corrosive at high temperatures, rendering crucibles unusable after 1 or 2 growths [12]. An extreme example, reflecting the high Pt solubility in the $SrCl_2$ melt, is the growth of $Sr_4PtO_6$ crystals of several mm in length from Pt crucibles by heating only hydrated $SrCl_2$ to 1150°C [13].

Significant variations in the magnetic and transport properties of $Sr_2IrO_4$ exist among previous reports, as well as in those of its *n*>1 sister compounds, assumed to reflect chemical disorder in the samples [9]. In particular, non-stoichiometric $Sr_{0.94}Ir_{0.78}O_x$ single crystals grown by using a soak temperature of 1470°C [10] were reported to exhibit very different magnetic properties from stoichiometric samples [5].

Conflicting crystal growth conditions for $Sr_3Ir_2O_7$ have been reported. In [8], single phase $Sr_3Ir_2O_7$ crystals could be prepared only in a narrow temperature range between 1480°C and 1440 °C, with a rapid quench to room temperature from 1440 °C in order to avoid the formation of a 214 impurity phase. Other authors report crystals $Sr_3Ir_2O_7$ growing upon cooling from 1440°C [14]. In yet another report, single-phased $Sr_3Ir_2O_7$ crystals were found to grow only below 1050°C, with an intergrowth of 214 and 327 phases occurring at an intermediate temperature range between 1150 °C and 1050 °C [9].

In the present work, we report a detailed and systematic study of the growth conditions for 214, 327 and 113 phases using a $SrCl_2$ flux well *below* its boiling point. This allowed us to use Pt crucibles over 10-20 times without any noticeable damage and to limit corrosion of heating and thermoelements of our furnaces. We show that the yield of the different phases strongly depends on the nutrient/solvent ratio for fixed soak temperature and cooling rate. Single crystals of $Sr_2IrO_4$, $Sr_3Ir_2O_7$ and $SrIrO_3$ have been grown in a reproducible way using identified and reproducible quasi-equilibrium growth conditions. All samples were characterized by means of energy dispersive x-ray spectroscopy, powder x-ray diffraction and magnetization measurements.

**Experimental details**

In all our flux-growth experiments we have used as a nutrient material a mixture of $SrCO_3$ (4N) and $IrO_2$ (4N) powders at 2:1 molar ratio, so that ~1.2 g of this mixture would yield 1 g of the 214 phase. As a solvent, we used anhydrous $SrCl_2$. Please note that the use of anyhydrous flux is a rather important experimental condition here [15]. The nutrient-to-flux weight ratio was varied from 1.2:5 to 1.2:20. All crystals were grown in a 50 ml Pt crucible covered with a Pt lid. The edges of the lid and the crucible were folded and tightly squeezed together in order to reduce flux evaporation. All growth experiments have been done in a low temperature gradient box furnace in air. The temperature profile used is depicted in Figure 1(a). Note that, a short dwell time is induced while the furnace warm-up rate is changed, so as to ensure that the furnace temperature doesn't overshoot from 1210 ºC. Crystals were detached from the solidified melt by washing with warm water. Typical crystals obtained by this procedure using different nutrient to solvent rations are shown in Figure 1(b)-(e). The chemical composition of the crystals was investigated by dispersive x-ray spectroscopy in a Zeiss EVO/MA 15 scanning electron microscope. The XRD patterns were collected in transmission geometry with Co and Cu K$\alpha_1$ radiation (used wavelength is mentioned in the associated figure) using a StoeStadi-Powder diffractometer equipped with a Ge(111) primary monochromator and a DectrisMythen 1 K detector. Magnetization of the crystals was measured between 2 and 300 K in a Quantum Design Magnetic Properties Measurement System with a SQUID detector.

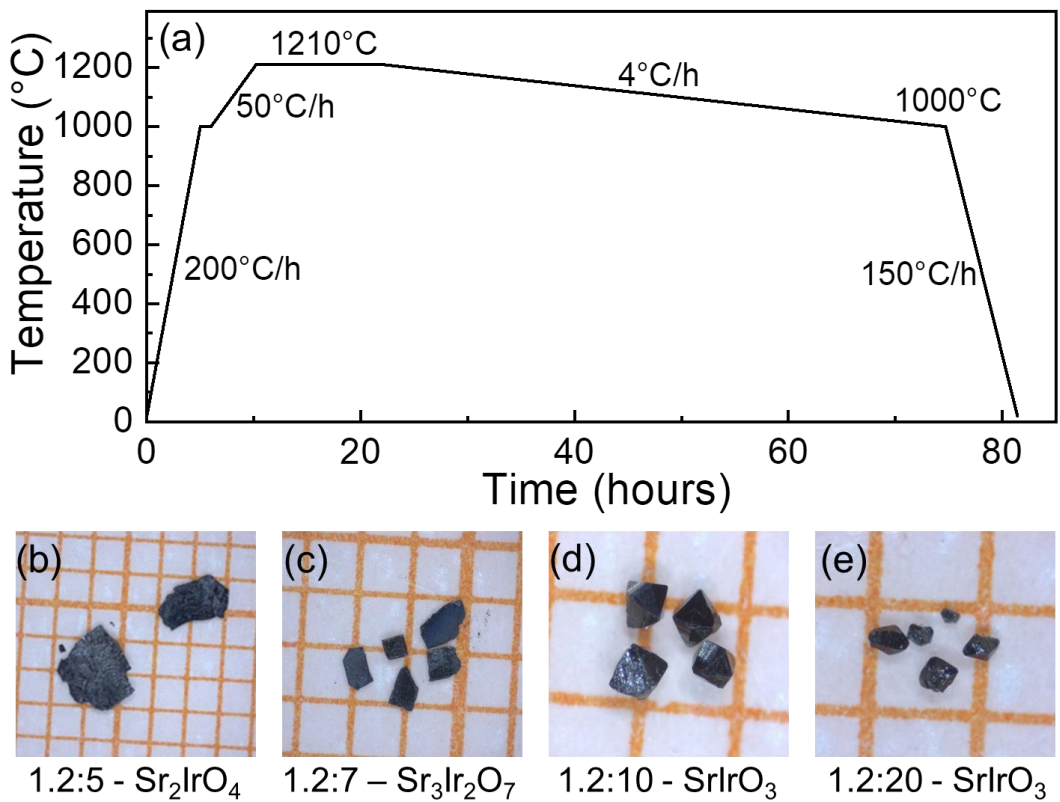

**Figure 1.** (a) Temperature profile used for crystal growth. Typical crystals obtained for nutrient: solvent weight ratios (obtained crystal) of (a) 1.2:5 (214), (b) 1.2:7 (327), (c) 1.2: 10 (113), (d) 1.2:20 (113). The underlying orange grid has a 1 mm spacing.

**Results and discussion**

As mentioned above, growing iridate crystals in boiling $SrCl_2$ provides very unstable conditions for crystal growth and can significantly shorten the lifetime of expensive Pt crucibles and lab furnaces. In order to avoid these problems, we used a soaking temperature of 1210°C, 40°C below the boiling point of anhydrous $SrCl_2$. Notice that our Pt crucibles were used in, at least, 10-20 runs continuously without any damage.

Single crystals of the 214 phase are obtained following the temperature vs. time profile described in Figure 1(a) and a nutrient-to-flux weight ratio of 1.2:5, as shown in Figure 1(c). Powder XRD measurements on crushed single crystals confirm the phase purity of our crystals, with *I*4$_1$/*acd* tetragonal lattice parameters *a*=5.497 Å and *c*=25.80 Å in good agreement with previous reports [16], see Figure 2. In contrast to other reports of crystals grown using soak temperatures of 1150°C or lower [9], we found no indication of the presence of a Sr$_3$Ir$_2$O$_7$ impurity phase in our 214 crystals, neither in XRD measurements nor in susceptibility curves (Figure 3, see discussion below). Since 327 undergoes a canted antiferromagnetic order at 285 K, intergrowth of 327 in 214 usually leads to a two-step structure in the magnetization curve [9] which is clearly absent in our data. The high chemical homogeneity of our 214 samples is further attested by the lack of contrast in backscattered electron images, see Figure 2(d). EDX measurements on several spots of the crystal, showed Sr:Ir ratios consistent with the expected 2:1 within the standard EDX accuracy. No evidence of Pt contamination was found. We compare the magnetization data for our 214 crystals with that of crystals grown after soaking at 1480 °C (for 20 h) [17] and 1300 °C (no dwell time) [9] in Figure 3. The transition temperature is determined from the maxima of the first order derivative as shown in the inset of Figure 3. Remarkably, samples prepared with the highest soaking temperature exhibit the lowest ordering temperature, while crystals prepared in this work and in [9] have very similar $T_N$ ~ 240 K. It is worth to mention that increasing the dwell time at high temperatures is also detrimental to $T_N$: only 214 crystals prepared without a dwell time from 1300°C presented $T_N$ ~ 240 K [9], while a 24 h dwell time at this temperature demonstrably decreased $T_N$ [9]. The origin of the variations in shape of the magnetization curves is unclear. Since anisotropy is large, it could be partly related to a small sample misorientation. In addition, the magnetic behavior is sensitive to subtle structural distortions. This requires a dedicated study beyond the scope of this work.

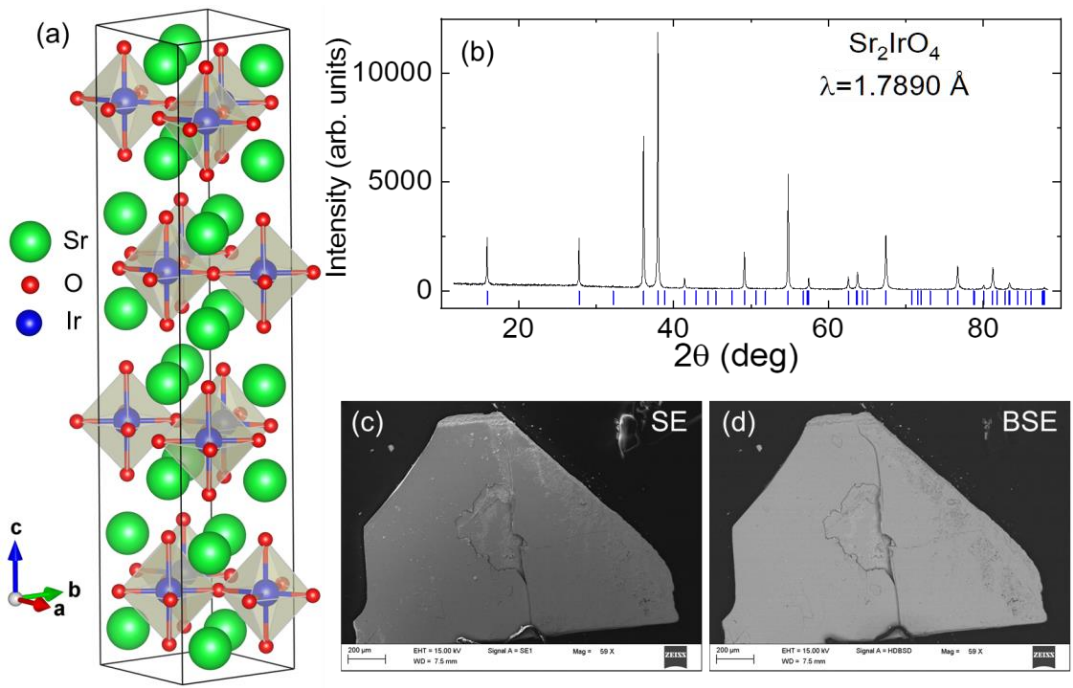

**Figure 2**. (a) Tetragonal crystal structure of $Sr_2IrO_4$. (b) Powder XRD measurement on crushed 214 crystals. Vertical bars correspond to the calculated Bragg peak positions for the 214 phase. (c) Secondary electron (SE) image of a crystal acquired with the scanning electron microscope. (d) Backscattered electron image (BSE) corresponding to (c) showing chemical homogeneity of the crystal.

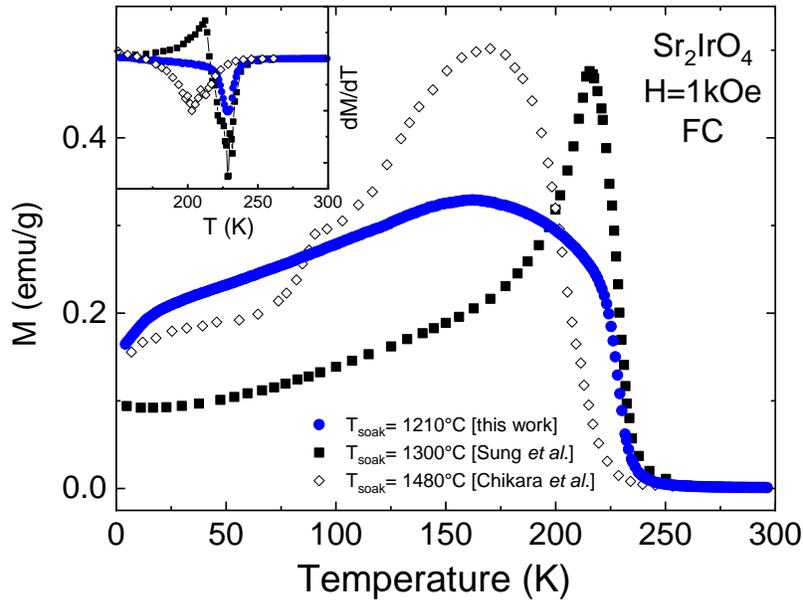

**Figure 3.** In-plane susceptibility measured after field cooling in 1 kOe for $Sr_2IrO_4$ crystals grown using different soaking temperatures. Chikara *et al.* heated the precursors to 1480°C, fired for 20h, cooled with 5°C/h to 1400°C and quenched to room temperature [8, 17]. Sung *et al.* fired the precursors at 1300°C, followed by cooling with 8°C/h to 900°C and quenching to room temperature [9]. Inset: derivative of the susceptibility as a function of temperature. The transition temperature is defined as the temperature at which the derivative of susceptibility has a minimum.

Using the temperature vs. time profile depicted on Figure 1(a) and a nutrient-to-flux weight ratio of 1.2:7, we obtained crystals of the 327 phase, as shown in Figure 1(c). Powder x-ray diffraction, shown in Figure 4, confirms the purity of the phase. The refined data are in good agreement with the reported $I4/mmm$ tetragonal structure, with lattice parameters $a = 3.899$ Å and $c = 20.894$ Å being very similar to previous reports [18]. EDX analysis yields a Sr:Ir ratio of 1.5, within the standard EDX accuracy, as expected for the 327 phase. The shiny white areas such as that indicated by the arrow in Figs 4(b) and (c) present a very high (70 at%) Pt content and are possibly small flakes of metallic Pt. The presence of such flakes is indicative of corrosion of the Pt crucible, in spite of the lower soaking temperatures used in this work. This finding suggests that growth procedures performed at higher temperatures may suffer from even more severe Pt corrosion. The effect of the Pt solubility in the melt will be reported in detail elsewhere [19].

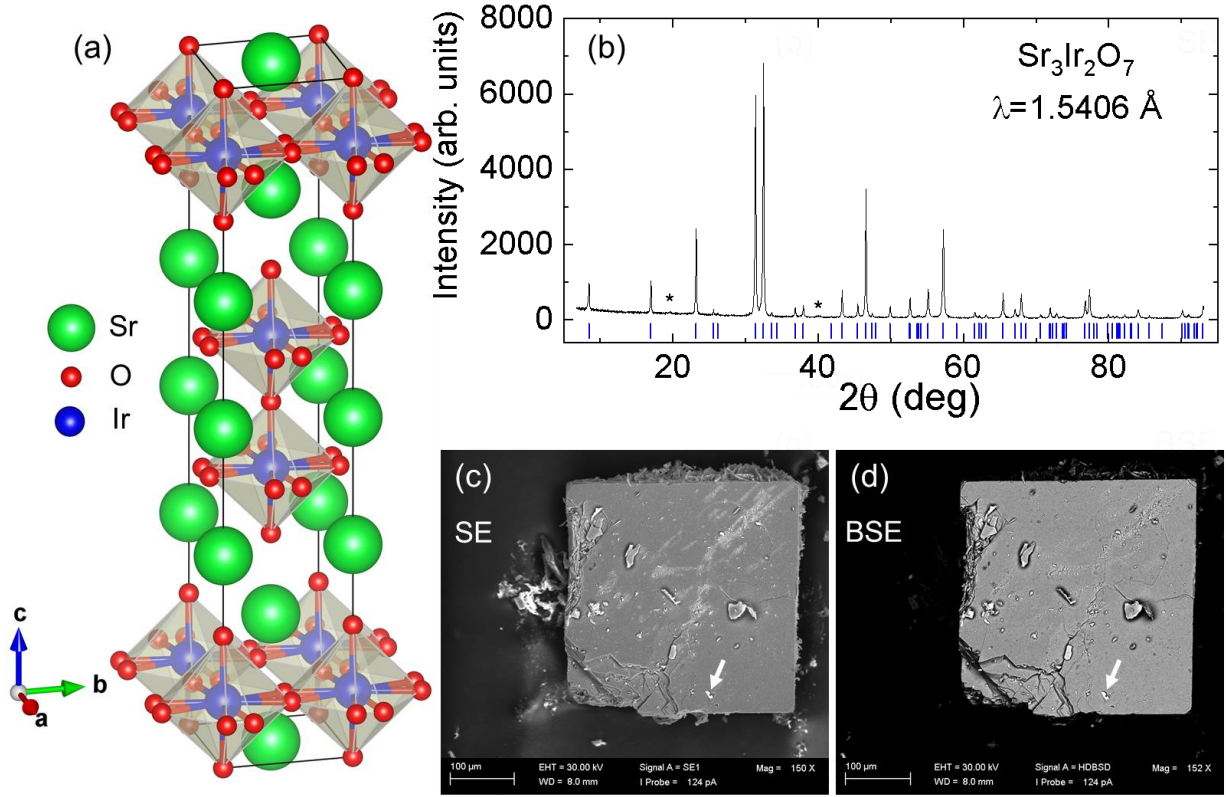

**Figure 4.** (a) Tetragonal crystal structure of $Sr_3Ir_2O_7$. (b) Powder diffraction pattern on crushed 327 single crystals. The vertical bars correspond to expected Bragg peak positions for the 327 phase. * represents impurity peaks from non-magnetic $Sr(OH)_2$, which might have formed during flux washing from unreacted SrO. (c) Secondary electron image of a typical 327 crystal. (d) Backscattered electron image corresponding to (c). Small bright flakes such as indicated by the arrow are mostly composed of Pt.

Magnetization curves for 327 crystals collected at a field of 1 kOe along the basal plane are presented in Figure 5. Our sample shows a magnetic transition around 285 K, with a steep downturn below 50 K, consistent with previous reports [8, 9]. The slight anomaly observed near 240 K in the ZFC magnetization probably indicates the presence of a tiny amount of the 214 phase, although the 214 phase was not detected in the powder diffraction measurement. Our 327 sample presents lower magnetization values in the paramagnetic phase than in previous reports. This could be related to a misalignment of the sample in the magnetic field, since 327 exhibits a considerable magnetic anisotropy, with lower susceptibility for fields applied along *c* [8]. Another possible explanation is that our crystals possess lower amount of magnetic defects usually associated with a chemical disorder.

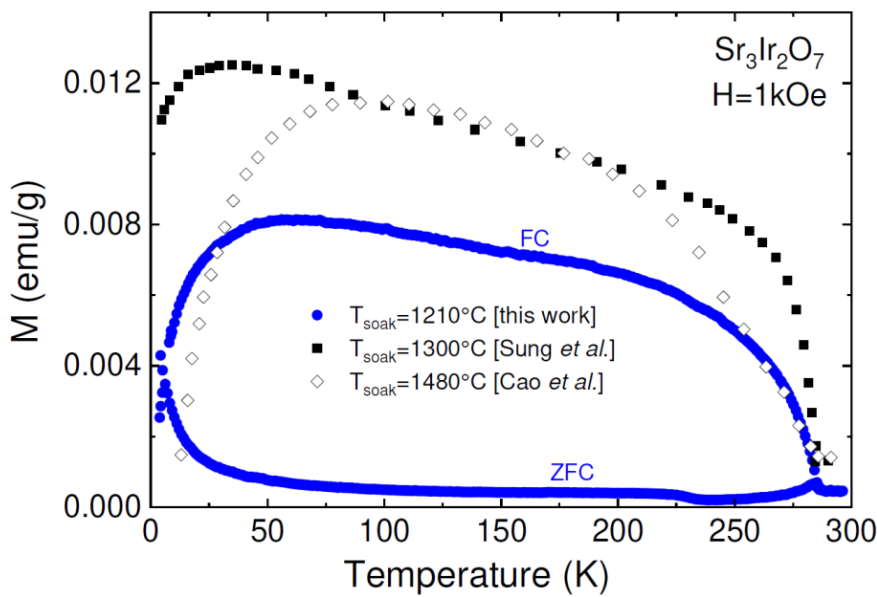

**Figure 5.** Magnetization of 327 crystals grown with different soaking temperatures [9,7] for magnetic field applied along the basal plane. The anomaly at ~235 K in the zero-field-cooled curve indicates the presence of a small amount 214 impurity phase.

Single crystals of the 113 phase could be grown for nutrient:flux weight ratios varying from 1.2:10 to 1.2:20 using the temperature-time profile shown on Figure 1. A ratio of 1.2:10 produced remarkably larger crystals than a 1.2:20 ratio, see Figures 1(d) and (e). Notice that the characteristic morphology of the crystals in Figure 6 is very different to that of layered 327 and 214 crystals (but with tetragonal symmetry lattice) reflecting the considerably different structure of the 113 compound (monoclinic). The Sr and Ir content determined by EDX analysis was found to be in agreement with the expected 1:1 atomic ratio, within the standard EDX accuracy. Occasionally, small peaks at the characteristic energies for Pt were detected by microprobe analysis, although too weak for a reliable quantification ($\lesssim$1%at). Whether this Pt is present only on the surface or incorporated in the bulk is currently unclear, since crystals are too small to be polished prior to EDX analysis. The powder XRD pattern, shown on Fig. 6, is in excellent agreement with the reported C12/c1 monoclinic structure, with lattice parameters $a$ = 5.594 Å, $b$ = 9.594 Å, $c$ = 14.147 Å, $\beta$ = 93.25°. No impurity phases were detected according to our XRD analysis. Magnetization data in Fig.7 exhibit paramagnetic behavior, with no irreversibility between field-cooled and zero-field-cooled curves. No anomalies at the ordering temperatures of 214 ($T_N$ ~ 240 K) and 327 ($T_N$ ~ 285 K) have been detected. A fit of the susceptibility including a constant term and a Curie-Weiss term yields relatively low values for the temperature independent susceptibility $\chi_o = 1.6 \cdot 10^{-4}$ emu/mol Oe, the Curie-Weiss temperature $\theta$ = -1.8 (2) K and an effective moment $\mu_{eff}$ = 0.47 $\mu_B$/Ir. These values are in contrast to a previous report for crystals grown using $T_{soak}$ = 1480°C, which exhibit $\chi_o > 1 \cdot 10^{-3}$ emu/mol Oe and "effective moments and Curie-Weiss temperatures that are much too large to be physically meaningful" [5]. It is worth to mention that the effective moment $\mu_{eff}$ = 0.47 $\mu_B$/Ir of our 113 crystals is typical for many iridates, for example, compare with $\mu_{eff}$ = 0.5 $\mu_B$/Ir reported for $Sr_2IrO_4$ [20], and $\mu_{eff}$ = 0.69 for $Sr_3Ir_2O_7$ [7]. All of this hints to a higher quality of our 113 crystals grown below the boiling point of the flux.

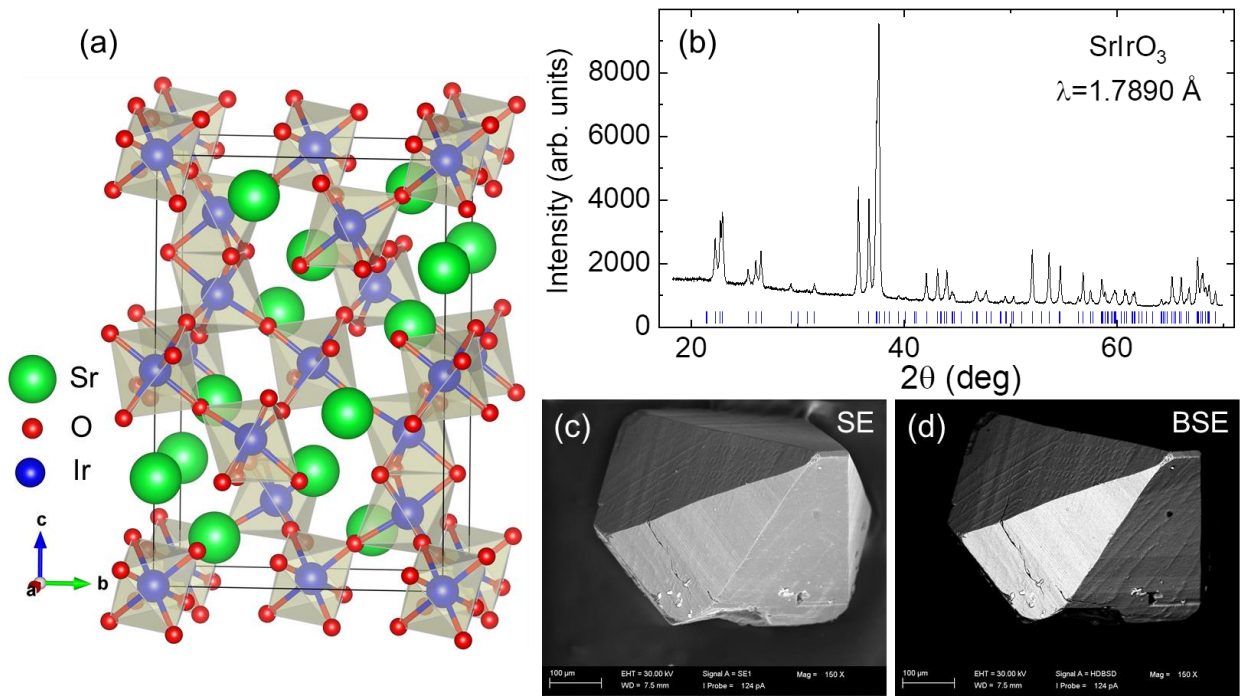

**Figure 6:** (a) Monoclinic crystal structure of SrIrO$_3$. (b) Powder diffraction pattern on crushed 113 single crystals. The position of SrIrO$_3$ Bragg peaks is indicated by vertical bars. (b) Secondary electron and (c) backscattered electron images of a typical 113 crystal, showing the characteristic morphology of this phase and chemical homogeneity.

In order to study the crystallization temperature range for the 113 phase, we considered an intermediate nutrient:solvent weight ratio of 1.2:15 and varied soaking temperature and cooling rate. Surprisingly, no 113 crystals were formed in a crystallization temperature range from 1210 °C to 1120 °C, even for a cooling rate of 2°C/h and a soak time of 20 h. Medium-sized crystals, typically ~500 μm long, were obtained in a crystallization temperature range from 1150 °C to 900 °C at 3 °C/h, while a lower 1050 - 900 °C temperature range yielded very tiny crystals (≤ 100 μm). This highlights the lower thermal stability of the 113 phase compared to 214 and 327 phases, and is consistent with a previous report indicating the decomposition of the 113 phase at 1205°C [6].

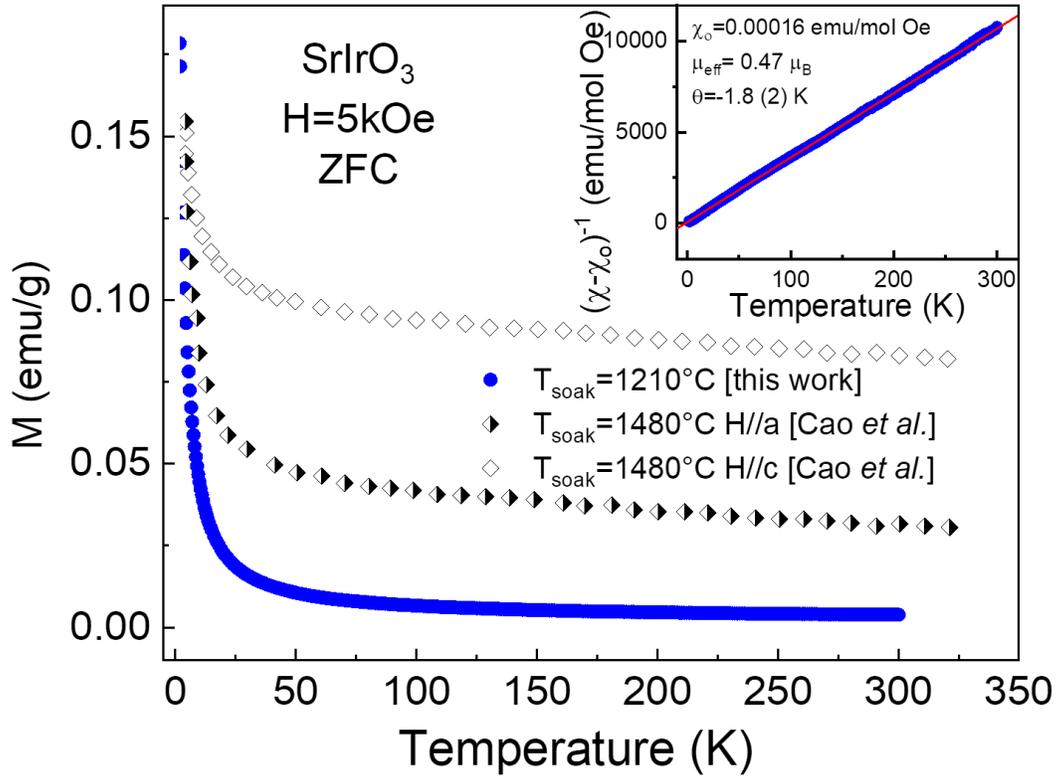

**Figure 7:** SrIrO$_3$ magnetization data. The data for crystals grown using T$_{soak}$ = 1480°C are taken from [5]. Inset: inverse susceptibility (blue circles) and Curie-Weiss fit (red line) for crystal grown using T$_{soak}$ = 1210°C.

**Conclusions**

We have shown that it is possible to selectively grow crystals of Sr$_{n+1}$Ir$_n$O$_{3n+1}$ with n = 1,2,∞ from anhydrous SrCl$_2$ flux below its boiling temperature, by using a 2:1 molar mixture of SrCO$_3$ and IrO$_2$ as a nutrient and a single temperature profile, but varying the nutrient-to-solvent ratio. For a nutrient-to-solvent weight ratio of 1.2:5, the 214 phase was crystallized, for 1.2:7 the 327 phase, while at the highest ratios between 1.2:10 and 1.2:20 the 113 phase was crystallized. The magnetic ordering temperatures of the 214 and 327 crystals are very similar to those reported in literature, and the 113 crystals exhibit a paramagnetic behavior down to 2 K similar to previous reports, albeit with a considerably smaller temperature independent susceptibility. The susceptibility of 113 crystals could be well fitted by a Curie-Weiss law including a *T*-independent contribution, in contrast to previous publications. Reducing the soak temperature below the solvent boiling point not only provides more stable and controllable growth conditions, but also extends considerably the lifetime of platinum crucibles and furnaces, thereby reducing growth costs.

## Acknowledgements

This work was partially supported by Deutsche Forschungsgemeinschaft (DFG) through the Collaborative Research Center SFB 1143 (Correlated Magnetism: from frustration to topology, project ID 247310070). The authors would like to thank S. Müller-Litvany for her help with EDX measurements, A. Omar and C. G. F. Blum with the XRD, A. U. B. Wolter and S. Gass for technical assistance with magnetization studies.